\documentclass[twoside]{report}
\usepackage{svcon2e}

\usepackage{amsmath}   
\usepackage{amssymb}
\usepackage{amsthm}

\newcommand{\ed}{\end{document}}
\renewcommand{\theequation}{\arabic{section}.\arabic{equation}}

\def\slapar{\partial\hspace { - .55 em}/}
\def\slaz{z\hspace { - .55 em}/}
\def\slac{C\hspace { - .55 em}/}
\def\g5{\gamma_5}
\def\gu{\gamma^\mu}
\def\smu{\sigma^{\mu\nu}}
\def\tmu{{\mathcal{T}}_{\mu\nu}}
\def\tnu{{\mathcal{T}}_{\nu\mu}}
\def\mB{{\mathcal{B}}}
\def\mS{{\mathcal{S}}}
\def\mV{{\mathcal{V}}}
\def\ty{{{\mathbf{t}}_Y}}
\def\text{\textrm}
\begin{document}
\chapter{Pauli Terms Must Be Absent\break In Dirac Equation}
\chapterauthors{Kurt Just and James Thevenot}
{{\renewcommand{\thefootnote}{\fnsymbol{footnote}}
\footnotetext{AMS Subject Classification: 81E10, 81Q05, 81S05.}
}}

\begin{abstract}
It should be of interest, whether Dirac's equation involves all 16 basis elements of his Clifford algebra $Cl_D.$ These include the 6 `tensorial' $\smu$ with which the `Pauli terms' are formed.
We find that these violate a basic axiom of any *-algebra, when Dirac's $\Psi$ is canonical.
Then the Dirac operator is spanned only by the 10 elements $1,i\gamma_5,\gamma^\mu,\gamma^\mu\gamma_5$ (which don't form a basis of $Cl_D$ because the
$\smu$ are excluded).\\
\noindent {\bf Keywords: }Quantum field theory, Dirac equation, Clifford algebra.\par
\end{abstract}

\pagestyle{myheadings}
\markboth{Kurt Just and James Thevenot}{Pauli Terms Must Be Absent In Dirac Equation}

\section{Motivation and conclusions}
\label{sec1}
In Dirac's equation
\begin{equation}
i\slapar\Psi = \mB \Psi \quad \text{with} \quad \slapar := \gu \frac{\partial}{\partial x^\mu}
\label{oneone}
\end{equation}
the Bose field
 $\mB$ is a member of the Clifford algebra $Cl_D.$  Hence it can be written as
\begin{equation}
\mB = \mS^+ + i \g5 \mS^- + \gu \mV^+_\mu + \gu\g5 \mV^-_\mu + \sigma^{\mu\nu} \tmu.
\label{onetwo}
\end{equation}
Here $ \mS^{\pm},  \mV^{\pm}, \tmu$ are matrices which act on the flavors and colors of $\Psi$ (the Dirac field for leptons and quarks).
In the excellently verified Standard Model,
the {\em matrices} 
\begin{equation}
\mS := \mS^+ + i \g5 \mS^- \quad \text{and} \quad 
\mV_\mu := \mV^+_\mu + \g5 \mV^-_\mu
\label{onethree}
\end{equation}
contain all the fields of Higgs and Yang-Mills.
Vice versa, the Standard Model requires (\ref{onetwo}) to contain the 10 basis elements $1,i\gamma_5,\gamma^\mu,\gamma^\mu\gamma_5 \in Cl_D,$ but not the further $\smu$ or $\smu \gamma_5$ (6 of which are linearly independent).
Thus we encounter the question, to what extent (\ref{onetwo}) can involve $\tmu = - \tnu.$
For this problem it is irrelevant whether $\tmu$ is a separate `tensor potential' or
a multiple of Maxwell's $F_{\mu\nu} := A_{\mu,\nu} - A_{\nu,\mu}$ (as proposed by Pauli \cite{refer1}) or of its dual $\varepsilon_{\mu\nu\rho\sigma}F^{\rho\sigma}.$ Hence we always call $\sigma^{\mu\nu}\tmu$ the `Pauli term' of (\ref{onetwo}). In rigorous, but not trivial ways,
we find that $\tmu$ must be {\em absent} for a very basic reason:
For the members 
$\mathbf{a},\mathbf{b},\ldots$ of any *-algebra and their conjugates
$\mathbf{a}^\dagger,\mathbf{b}^\dagger,\ldots,$ one postulates
$(\mathbf{a}^\dagger \mathbf{b})^\dagger = \mathbf{b}^\dagger \mathbf{a}.$
This would be violated by any Pauli term. Hence we must demand
\begin{equation}
\tmu = 0 \quad \quad \text{in order to keep} \quad \quad
(\mathbf{a}^\dagger \mathbf{b})^\dagger = \mathbf{b}^\dagger \mathbf{a}.
\label{onefour}
\end{equation}
In other words, our fields will not generate a *-algebra, unless (\ref{onetwo}) is 
{\em restricted} by (\ref{onefour}). This clear result has not been found in the literature,
because a familiar reciprocity \cite{refer2} is generally misnamed a theorem, whereas we prove it to be a {\em condition}, which excludes (\ref{onetwo}) unless it satisfies (\ref{onefour}).

Showing in Sections \ref{sec2} and \ref{sec3} the adopted foundations, we indicate the proof of (\ref{onefour})
briefly in Section \ref{sec4}, and more elaborately in Appendix \ref{appA}. It rests on a reciprocity {\em condition}, which is discussed in Appendix \ref{appB}.  Calling this a `relation' \cite{refer2}, one generally suggests that it holds  without restricting the Bose fields to be prescribed.  That misnomer may have caused the absence of 
(\ref{onefour}) in the literature.

\section{Gauge theory from Clifford algebra}
\label{sec2}
For the vector field from (\ref{onethree}) we must admit the gauge transformation
\begin{equation}
\mV_\mu \quad \rightarrow \quad e^{-i\omega}\left(\mV_\mu - i \partial_\mu\right)e^{i\omega}
\quad
\approx
\quad
\mV_\mu + \omega_{,\mu} + i \left[\mV_\mu,\omega\right].
\label{twoone}
\end{equation}
In order to state that $\mV_\mu$ and $\omega$ are {\em hermitian} matrices, we write them
\begin{equation}
\mV_\mu(x) = \ty \mV^Y_\mu(x) = \mV_\mu(x)^\dagger
\quad
\text{and}
\quad
\omega(x) = \ty \omega^Y(x) = \omega(x)^\dagger.
\label{twotwo}
\end{equation}
While the constant matrices $\ty$ act on flavors and colors, they contain all coupling constants and the $\gamma_5 := i \gamma^0 \gamma^1 \gamma^2 \gamma^3$ from Dirac's Clifford algebra $Cl_D.$
We generate this $Cl_D$ by $\gamma^{(\mu}\gamma^{\rho)} = \eta^{\mu\rho}$ from
\begin{equation}
\gamma^\dagger_\mu = \gamma^{-1}_\mu = \gamma^\mu = \overline{\gamma^\mu}
\quad,
\quad
\text{where}
\quad
\overline{\Gamma} := \gamma_0 \Gamma^\dagger \gamma_0 \in Cl_D.
\label{twothree}
\end{equation}
A transformation similar to the homogeneous part of (\ref{twoone}) follows for the $\mS$ of (\ref{onethree}). In order to prove (\ref{onefour}), however, we must initially use  (\ref{onetwo}) with (\ref{onethree}) in the form
\begin{equation}
{
\mathcal{B} = \mS + \gamma^\mu \mV_\mu + \smu \tmu,
\quad 
\text{where}
\quad
\tmu \not=}\ 0.
\label{twofour}
\end{equation}
This together with $(\mathbf{a}^\dagger \mathbf{b})^\dagger = \mathbf{b}^\dagger \mathbf{a}$
will in Section \ref{sec4} yield a contradiction, which then proves (\ref{onefour}).  
That proof holds in Quantum Induction \cite{refer3} where the canonical relations
\begin{equation}
{\left[ \Psi(x),\Psi(0)^\dagger\right]}_+ \delta(x^0) = \delta(x)
\quad
\text{and}
\quad
{\left[ \Psi(x),\Psi(0)^T\right]}_+ \delta(x^0) = 0
\label{twofive}
\end{equation}
together with Dirac's equation (\ref{oneone}) are {\em fundamental}. Its possible validity under presumptions different from (\ref{oneone}) through (\ref{onethree}) is discussed in 
Appendix \ref{appB}.

\section{Short distance representation}
\label{sec3}

For the bilocal, time ordered Dirac matrix
\begin{equation}
b(x,z) := (4\pi)^2 \, T \, \Psi(x + z) \overline{\Psi}(x - z),
\label{twosix}
\end{equation}
(\ref{oneone}) and (\ref{twofive}) provide the differential equation 
\begin{equation}
\left\{\slapar^x + \slapar^z + 2i \mB(x+z) \right\}
b(x,z) = 2 \pi^2 \delta(z) = i \slapar^z \slaz^{-3}_- ,
\label{twoseven}
\end{equation}
where $ \slaz^{-3}_- := (z^2 - i\epsilon)^{-2}\slaz $
with
$\epsilon \rightarrow +0.$            
The representation used here for $\delta(z) := \delta(z^0) \delta(z^1) \delta(z^2) \delta(z^3) $
follows directly from the familiar
$$\square\,(z^2 - i\epsilon)^{-1} = (2\pi)^2 i \delta(z).$$          
Writing (\ref{twosix}) as
\begin{equation}
b(x,z) = i \slaz^{-3}_- + (C^{-2} + C^{-1} + r^0)(x,z),
\label{threeone}
\end{equation}  
let us anticipate that the $C^h$ can be made {\em homogeneous} in the sense that
\begin{equation}
C^h(x,\lambda z) = \lambda^h C^h(x,z)
\quad
\text{for}
\quad
h = -2, -1
\quad
\text{and}
\quad
\lambda \in \slac.
\label{threetwo}
\end{equation}
Also using the `Taylor representation'
\begin{equation}
\mB(x+z) = \mB(x) + z^\mu\mB_{,\mu}(x) + R(x,z) \slaz, \quad
\text{where} \quad R(x,0) = 0,
\label{threethree}
\end{equation}
we can split (\ref{twoseven}) into
\begin{gather}
\slapar^z C^{-2}(x,z) = 2 \mB(x) \slaz^{-3}_- , \label{threefour} \\
\slapar^z C^{-1}(x,z) = 2 z^\mu \mB(x)_{,\mu} \slaz^{-3}_-
                        - \{ 2 i \mB(x) + \slapar^x \} C^{-2}(x,z), \label{threefive} \\
\slapar^z r^0(x,z) = \ldots  - 2 i R(x,z) z^{-2}_- .
\label{threesix}
\end{gather}
Here the dots symbolize infinitely many unknown terms; they will be irrelevant
because they are not more singular than $\slaz^{-1}_-$ (for $z \rightarrow 0$ at
$z^2 \not= 0)$.
As one easily verifies, (\ref{threefour}) with (\ref{twofour}) can be solved by
\begin{equation}
z^4 C^{-2}(x,z) = 2 \slaz z^\mu \overline{\mV}_\mu(x)
- z^2 \mS(x)^\dagger + \slaz \sigma^{\mu\nu} \slaz  \tmu(x)^\dagger.
\label{threeseven}
\end{equation}
This clearly satisfies (\ref{threetwo}) and due to Appendix \ref{appA} is the {\em only} solution of (\ref{threefour}) which does so. Since it makes the right side of (\ref{threefive})
homogeneous in $z$ of the order $h = - 2,$ we can choose also $C^{-1}(x,z)$ to obey
(\ref{threetwo}). On the right side of (\ref{threesix}), we have omitted terms which due to (\ref{threeseven}) are as homogeneous as $\slaz^{-1}_-$ (or less singular).

Thus (\ref{threesix}) can be solved by an $r^0$ which (due to $R(x,0) = 0)$ satisfies
\begin{equation}
z^\mu r^0(x,z) \rightarrow 0
\quad
\text{for}
\quad
z \rightarrow 0.
\label{threeeight}
\end{equation}
Therefore, we can include in $r^0$ those terms from $C^{-2}$ and $C^{-1}$ which are left
arbitrary by (\ref{threefour}) and (\ref{threefive}) because they are independent of $z.$
In order to learn much more about $r^0(x,z)$ than (\ref{threeeight}) shows, one would need `outer' boundary conditions (at large $z).$ These could be obtained from heat kernels (HK);
but here the `inner' condition (by (\ref{twofive}) giving to (\ref{twoseven}) its right
side) has been sufficient.

No obstruction to {\em solving} (\ref{twoseven}) has been encountered in (\ref{threeseven}) or in Appendix \ref{appA}.  Neither would any arise if we would extend our recursion for (\ref{threeone}) or invoke HK\cite{refer5}. Taking the Dirac adjoint of (\ref{twosix}),
however, we obtain the `reciprocity' condition 
\begin{equation}
\overline{z^4 b(x,z)} = z^4 b(x,-z) 
\quad
\text{due to}
\quad
(\Psi_+\Psi^\dagger_-)^\dagger =
\Psi_-\Psi^\dagger_+ .
\label{threenine}
\end{equation}
The latter exemplifies a general rule for *-algebras.

\section{Reciprocity as a condition}
\label{sec4}
The $z^4$ in (\ref{threenine}) removes the denominators of the terms in (\ref{threeone}), so that
the hermitian conjugation affects only their {\em numerators}. Hence these must
(for $h=-2,-1)$ satisfy        
\begin{equation}
\overline{z^4C^h(x,z)} = z^4 C^h(x,-z)
\quad
\text{and}
\quad
\overline{z^4r^0(x,z)} = z^4 r^0(x,-z),
\label{fourone}
\end{equation}
because all three are linearly independent. In (\ref{fourone}) we did not mention the
leading $i \slaz^{-3}_-$ of (\ref{threeone}), because it satisfies (\ref{threenine})
trivially.  Since (\ref{twofour}) equals its adjoint $ \overline{{\mathcal{B}}} = \overline{\mS} + \gu \mV^\dagger_\mu + \sigma^{\mu\nu} \overline{\tmu},$
that reciprocity is also verified for (\ref{threeseven}).

Instead of (\ref{fourone}) with $h=-1,$ however, in Appendix \ref{appA} we find
\begin{equation}
\overline{z^4C^{-1}(x,z)} \not= z^4 C^{-1}(x,-z)
\quad
\quad
\text{unless}
\quad
\quad
\tmu = 0.
\label{fourthree}
\end{equation}
Hence it is {\em misleading} when one calls (\ref{threenine}) a reciprocity `theorem' \cite{refer2}, as if it were fulfilled for arbitrarily prescribed Bose fields (\ref{twofour}).
Since (\ref{threeseven}) satisfies (\ref{fourone}) even with $\tmu \not= 0,$ we also see that (\ref{onefour}) cannot be derived as long as one only examines that solution of (\ref{threefour}). In other words, we do not know a way of reaching the conclusion (\ref{onefour}) without noting that the solution of (\ref{threefive}) provides (\ref{fourthree}).

By (\ref{onefour}), however, the $C^{-1}(x,z)$ solving (\ref{threefive}) is simplified greatly;  and the further parts of (\ref{threeone}) are {\em shortened} still more drastically.
Hence superfluous work is done by those who pursue higher terms of (\ref{threeone}) without inserting (\ref{onefour}) quickly. They solve the differential equation (\ref{twoseven}) correctly but in excessive generality. Thus they miss the fact that (\ref{twosix}) must also satisfy (\ref{threenine}), which is a non-trivial condition not expressed by (\ref{twoseven}).

\section*{Acknowledgments}  For comments we are thankful to S. A. Fulling, K. Kwong,
Z. Oziewicz, W. Stoeger and E. Sucipto.    

\renewcommand{\thesection}{\Alph{section}}
\setcounter{section}{0}
\renewcommand{\theequation}{\Alph{section}.\arabic{equation}}
\section{Appendix: Linear Differential Equations}
\label{appA}
\subsection{Exactly homogeneous solutions}
\label{appA1}

In (\ref{twoseven}) we have used
\begin{equation}
i\slapar^z \slaz^{-3}_- = 2 \pi^2 \delta(z)
\quad
\text{for}
\quad
\slaz^{-3}_- :=
(z^{-2}_-)^2 \slaz,
\label{a1}
\end{equation}
where $z^{-2}_- = (z^2 -i\epsilon)^{-1}$ with $\epsilon \rightarrow +0.$ These obviously provide
\begin{gather}
\slapar^z \slaz^{-3}_- z^\mu = \gamma ^\mu \slaz^{-3}_-, 
\quad
\slapar^z z^{-2}_- = - 2 \slaz^{-3}_-, \notag \\
\slapar^z \slaz^{-3}_- \smu \slaz = \gamma^\rho \slaz^{-3}_- \smu \gamma_\rho
= 2 \smu \slaz^{-3}_-,
\label{a2}
\end{gather}
which are derived most {\em easily}, when one uses (\ref{a1}) as often as possible. Thus (\ref{threeseven}) makes 
\begin{equation}
\slapar^zC^{-2}(x,z) =
2 \gu \slaz^{-3}_- \overline{\mV}_\mu(z) + 2 \slaz^{-3}_- \mS(x)^\dagger
+2 \slaz^{-3}_- \sigma^{\mu\nu} \tmu(x)^\dagger,
\label{a3}
\end{equation}
so that (\ref{threefour}) with (\ref{twofour}) is fulfilled.

All these calculations of course do not make sense on the cone $z^2 = 0.$ Hence throughout this paper  we assume $z^2 \not = 0$ , as we must clearly do in (\ref{threeone}) through
(\ref{threesix}). This is also true for the {\em limits} with $z \rightarrow 0,$
as needed in (\ref{threeeight}) and in the proof of (\ref{a1}). Hence such limit transitions can proceed on any path which ends at $z=0,$ except that it must not touch the cone $z^2 = 0.$

While (\ref{threefour}) with (\ref{twofour}) is due to (\ref{a1}) satisfied by
(\ref{threeseven}), its most general solution follows when we add any matrix $H$ which fulfills the homogeneous Dirac equation $\slapar H = 0.$ For choosing this $H$  we need the

\vspace{6 pt}
\noindent
{\bf Singularity Theorem}: Every Poincar\'{e} covariant member $H$ of Dirac's Clifford algebra
$Cl_D,$ which solves
\begin{equation}
\slapar H(z) = 0 
\quad
\text{in a neighborhood of}
\quad
z = 0,
\label{a4}
\end{equation}
becomes for $z \rightarrow 0$ with $z^2 \not = 0 $ either more singular than $\slaz^{-3}$ or less than $\slaz^{-1},$ hence yields either 
\begin{equation}
\lim_{z\rightarrow 0} \slaz^3 H(z) = \infty
\quad
\text{or}
\quad
\lim_{z\rightarrow0} z^\mu H(z) = 0.
\label{a5}
\end{equation}
One proves this easily when $H$ depends only on $z.$  The general proof is lengthy and scarcely of interest to physicists; hence we shall show it whenever requested.

Obvious solutions of (\ref{a4}) and (\ref{a5}) are all $H$ which do not depend on $z.$  Already the solution of (\ref{threefour}) by (\ref{threeseven}) says that (\ref{threetwo}) with 
$h = -2$ can be satisfied. The theorem (\ref{a5}) proves that (\ref{threeseven}) yields the only
$C^{-2}(x,z)$ which does so (such that (\ref{threefour}) and (\ref{threetwo}) make
(\ref{threeseven}) necessary).  It also shows that the leading term of (\ref{threeone}) is determined uniquely. In the same way, (\ref{a5}) says that the solution of (\ref{threefive}) and (\ref{threetwo}) will be {\em unique} when such a $C^{-1}(x,z)$ can be found at all.

\subsection{Relevant short distance terms}
\label{appA2}
Inserting (\ref{twofour}) and (\ref{threeseven}) in (\ref{threefive}), we obtain
\begin{align}
\slapar^zC^{-1}(x,z) 
& = 2 z^\mu \gamma^\rho \slaz^{-3}_- (\overline{\mV}_{\rho,\mu}
-\overline{\mV}_{\mu,\rho} - 2 i \overline{\mV}_\rho \overline{\mV}_\mu) 
\notag \\ 
& \phantom{=} + z^{-2}_- \gamma^\rho (\mS^\dagger_{,\rho} + 2 i \mV_\rho \mS^\dagger)  
+ 2 \slaz^{-3}_- z^\rho (\mS^\dagger_{,\rho}- 2 i \mS^\dagger \overline{\mV}_\rho) 
\notag \\ 
& \phantom{=} + 2 i z^{-2}_- \mS \mS^\dagger + 2 \slaz^{-3}_- \gamma^{\mu\nu} \slaz \mS \tmu 
\notag \\ 
& \phantom{=} + 2 z^\rho \gamma^{\mu\nu} \slaz^{-3}_- ( 2 \tmu \overline{\mV}_\rho + i{\mathcal{T}}_{\mu\nu,\rho}) - 2 z^{-2}_- \gamma^{\mu\nu} \tmu \mS^\dagger
\notag \\ 
& \phantom{=} + 2 i \gamma^{\mu\nu} \slaz^{-3}_- \gamma^{\rho\sigma} \slaz \tmu {\mathcal{T}}_{\rho\sigma} 
\notag \\
& \phantom{=} + \gamma^\rho \slaz^{-3}_- \gamma^{\mu\nu} \slaz (2 \mV_\rho \tmu -
i {\mathcal{T}}_{\mu\nu,\rho}) 
\label{a6}
\end{align}  
with
$\gamma^{\mu\nu} := \gamma^{[\mu}\gamma^{\nu]} = - i \sigma^{\mu\nu} = 
\gamma^\mu \gamma^\nu - \eta^{\mu\nu}.$ Here all the fields $\mS, \mV_\mu, \tmu$ and
their partial derivatives are localized at $x$; hence this parameter has been suppressed in the notation. Although (\ref{a6}) is for $C^{-1}$ a linear differential equation in $z$ of the {\em first} order, deriving a solution was tedious; but after such a $C^{-1}$ has been found,
only differentiations are needed to verify its validity. Only a few readers, however, would perform this extremely easy but time-consuming task; thus let us merely say that the $C^{-1}(x,z)$ solving the differential equation (\ref{a6}) is twice as lengthy as this.

It implies (\ref{fourthree}) and thus contradicts (\ref{fourone}), {\em unless} all 5 local field polynomials
\begin{align}
{\mathcal{Z}}^1_{\rho\sigma\tau\mu} &:= 
\left[
 {\mathcal{T}}_{[\rho\sigma} , 
 {\mathcal{T}}_{\tau]\mu}\right]_+ , 
\notag \\
{\mathcal{Z}}^2_{\rho\sigma\tau} &:=
{\mathcal{T}}_{[\rho\sigma} \overline{\mV}_{\tau]} + \mV_{[\tau} {\mathcal{T}}_{\rho\sigma]}, 
\notag \\ 
{\mathcal{Z}}^3_{\rho} &:=
\left[
 \mV^\mu ,
 {\mathcal{T}}_{\mu\rho}\right] 
\label{twentynine} \\
{\mathcal{Z}}^4_{\rho\sigma} &:=
{\mathcal{T}}_{\rho\sigma}\mS^\dagger + \mS {\mathcal{T}}_{\rho\sigma}, 
\notag \\
{\mathcal{Z}}^5_{\rho\sigma\tau} &:=
{\mathcal{T}}_{\rho\sigma,\tau} + i \overline{\mV}_{\tau} {\mathcal{T}}_{\rho\sigma} -
i {\mathcal{T}}_{\rho\sigma} \mV_{\tau}
\notag
\end{align}
satisfy
\begin{equation}
{\mathcal{Z}}^n_{...}(x) = 0.
\label{thirty}
\end{equation}
These derivations have been rigorous, because we did not admit any approximation (rarely possible in physics).  Our solution (\ref{onefour}) is clearly the only which holds irrespective of $\mS$ and $\mV_\mu.$ If (\ref{thirty}) were mathematically solvable by any  ${\mathcal{T}}_{\mu\nu} \not= 0$ (a case we can't examine exactly), it would restrict $\mS$ and $\mV_\mu$ in complicated and extremely unphysical ways.

Within the permitted size of this paper, we can't show the complete proof for the necessity of (\ref{thirty}); the known extension to quantum fields \cite{refer3} would enlarge it enormously.
A result as simple as (\ref{onefour}), however, should find a much shorter proof.  Hence we would prefer to delay the publication until that simplicity is achieved. If we would not show the result (\ref{onefour}) and indicate our lengthy derivation from (\ref{threenine}), however, it would be hard to get anyone interested in such a problem.

\section{The Reciprocity Violation}
\label{appB}

Let us finally collect further remarks about our result (\ref{onefour}) and
its absence from the literature:

\newcounter{jidx}
\begin{list}
  {(\alph{jidx})}{\usecounter{jidx}
     \setlength{\rightmargin}{\leftmargin}}
\item The reciprocity condition (\ref{threenine}) has been called \cite{refer2}
a {\em relation}, as if it had been proved with (\ref{oneone}) containing the
most general Bose field (\ref{onetwo}).
\item We had to perform extensive {\em computer} algebra for the derivation of
$C^{-1}(x,z)$ from its differential equation (\ref{a6}).  Still more would be required if one
were to make the compact result from HK explicit.  Neither is needed any longer, because it is easy to verify a known
solution of any differential equation, no matter how hard its integration
had been.  Because of (a), however, nobody found this tedious search
worthwhile.
\item In the mathematics of HK, the boundary conditions \cite{refer5}
at {\em large} $z$ are presently more interesting than the
behavior of (\ref{threeone}) at small $z.$
\item  It is fashionable, instead of the differential equation (\ref{twoseven})
to solve a related integral equation, with an `inner' boundary condition
given by the right side of (\ref{twoseven}) and a purely 
mathematical condition at some outer boundary (where $z$ is large or infinite).
For our problem, no choice of the latter makes sense because the result
(\ref{thirty}) depends only on (\ref{twoseven}) and its boundary condition at
$z = 0.$ Why should we use a physically irrelevant integral equation for a
conclusion which is {\em completely} determined by a differential equation
together with a single, well justified boundary condition?
\item Methods of HK have been initiated \cite{refer6} for {\em classical}
field theories, where the reciprocity arises from a symmetry of their
Green functions.
\item In many treatments by HK, not only the Bose field $\mB$ but also Dirac's
$\Psi$ is non-quantized (or not even mentioned).
Then (\ref{twoseven}) is regarded as an equation for a classical Green
function $b(x,z),$ not related to any quantum field such as
(\ref{twosix}).  In that approach, one hardly sees whether a reciprocity
(\ref{threenine}) should be desired.
\item Under {\em infinite} renormalizations, (\ref{oneone}) and
(\ref{twofive}) and therefore (\ref{threeone}) break down \cite{refer7}.
Hence we can't prove (\ref{onefour}) in familiar settings (although it might
be true even there).
\item Any significant $\tmu \not = 0$ would damage the excellent
{\em verification} of the Standard Model \cite{refer8} by the
magnetic moment of the electron.  This agreement \cite{refer9a} had formerly been regarded as a
brilliant confirmation of renormalized QED.
Under the present philosophy \cite{refer9} of
`effective' actions, however, it is an unimportant result of imprecise
measurements.
\item Instead of (\ref{twosix}), one often uses
$\beta(e,t) := (4\pi)^2 \text{T} \Psi(e+t) 
\overline{\Psi}(e) = b(e + \frac12 t, \frac12 t)$
with the {\em eccentric} coordinates $e = x-z$ and $t = 2z.$
These simplify  (especially under gravity) the
derivation of (\ref{twoseven}), but make the analysis of
(\ref{threenine}) complicated.
\item The singularity at $z^2 = 0$ makes (\ref{threeone}) dependent on the
{\em time ordering} of (\ref{twosix}). Hence (\ref{threenine}) can't simply be written
$\overline{b}(x,z) = b(x,-z)$ because an hermitian conjugation reverses the
time order.
\item Wherever basic `tensor potentials' $\tmu$ have found any attention, one
has {\em coupled} them to each other or further Bose fields \cite{refer10},
leaving their interaction with Dirac's $\Psi$ open.
\item Whenever `tensor couplings' are mentioned in phenomenology \cite{refer11},
it is unclear whether they are {\em fundamental} or caused by bound states or
by `radiative' corrections.
\item The {\em differential} or integral equation for (\ref{twosix}) can
`mathematically' be solved \cite{refer5} without any concern about the
reciprocity (\ref{threenine}) needed in physics.
However, (\ref{twoseven}) without (\ref{threenine}) does not exhaust
the contents of (\ref{twosix}).
\item The leading terms $i \slaz^{-3}_-$ and
$C^{-2}(x,z)$ of (\ref{threeone}) satisfy (\ref{threenine}) even
when (\ref{twofour}) has $\tmu \not = 0.$ Hence the {\em contradiction} between
(\ref{fourone}) and (\ref{fourthree})
is not recognized until one also determines $C^{-1}(x,z).$
\item A further reason for the usual rejection of (\ref{onefour}) may be
that such a clear result deserves a {\em simple} derivation.  Instead our
proof has required the lengthy (but straightforward) deduction of the 
differential equation (\ref{a6}) and its explicit solution.
That $C^{-1}(x,z)$ is twice as long as (\ref{a6}) and therefore not shown
here; but we hope that others can simplify our arguments.
\item In order to examine (\ref{threenine}) completely , 
the reciprocity (\ref{fourone}) should also be checked for the
`remainder' $r^0(x,z).$
Analyzing those parts of it which in $z$ are homogenous of the
orders $h = 0$ and
$h = 1,$ we have not found any restriction beyond $\tmu = 0$ (which
simplifies those
parts enormously).  Our {\em local} approach (without outer
boundary conditions) cannot extend that result to all orders.  This
should be taken as an incentive to treat the condition (\ref{threenine})
globally, but not as excuse for discarding our result (\ref{fourthree}).
Doing so would correspond to ignoring the singular part of a Laurent
series until all its orders are known.
\item For many authors, the Higgs field does not contribute to $\mathcal{B},$
because they attach it in {\em isospinors} to $\Psi.$
\item Many authors prefer {\em two-component} spinors instead of Dirac's~$\Psi.$
\item Some authors use other {\em notations} for Dirac matrices, for instance
$\mathbf{\alpha},\beta$ instead of $\gamma^\mu$ or explicit 4 x 4 squares.
\item For Dirac's $\gamma^\mu,$ one sometimes uses {\em representations}
in which $\gamma^\mu$ is not its own adjoint or the $\beta$ in
$\overline{\Psi} = \Psi^\dagger \beta$ differs from $\gamma^0.$
\item From (\ref{twofour}) we derived (\ref{onefour}) by
`reductio ad absurdum',
which not every mathematician appreciates.
\item Readers may dislike (\ref{onefour}), because beautiful theories are no
longer expected to be simple but to offer rich mathematical structures.
\item Instead of the Dirac equation for physics (which is of first order in
 Minkowski space), mathematicians prefer the elliptic equation
given by its iteration in Euclidean space.
\item One often uses Dirac's Clifford algebra without any basis, hence not
separating the scalar, vectorial and tensorial parts of $\mB$.
\end{list}

\pagebreak


\vskip 1pc
{\obeylines
\noindent Kurt Just
\noindent Department of Physics
\noindent University of Arizona
\noindent Tucson AZ 85721
\noindent E-mail: just@physics.arizona.edu
\vskip 1pc
\noindent James Thevenot
\noindent Department of Physics
\noindent University of Arizona
\noindent Tucson AZ 85721
\noindent E-mail: jimthev@physics.arizona.edu
}
\vskip 6pt
\noindent Received: September 30, 1999; Revised: February 6, 2000.
\end{document}